\documentclass[preprint,showpacs,preprintnumbers,amsmath,amssymb,nofootinbib]{revtex4}

\newcommand{\vect}[1]{\mathbf{#1}}

\usepackage{epsfig,graphicx}
\usepackage{dcolumn}
\usepackage{bm}

\begin{document}

\preprint{APS/123-QED}

\title{ Solvent--mediated interactions between nanoparticles at fluid interfaces}

\author{Fernando Bresme}
  \email{f.bresme@imperial.ac.uk}
 \affiliation{Department of Chemistry, Imperial College London, SW7 2AZ, United Kingdom}
\author{Hartwig Lehle}
\affiliation{
Max--Planck--Institut f\"ur  Metallforschung, Heisenbergstr. 3, D-70569 Stuttgart and 
Robert Bosch GmbH, Wernerstrasse 51, D-70469 Stuttgart, Germany
}
\author{Martin Oettel}%
 \email{oettelm@uni-mainz.de}
\affiliation{%
Johannes--Gutenberg--Universit\"at Mainz, Institut f\"ur  Physik, WA 331, D-55099 Mainz, Germany
}%

\date{\today}

\begin{abstract}
We investigate the solvent mediated interactions between nanoparticles adsorbed at a liquid--vapor interface in comparison to the solvent mediated interactions in the bulk liquid and vapor phases of a Lennard--Jones solvent. Molecular dynamics simulation data for the latter are in good agreement with results from integral equations in the reference functional approximation and a simple geometric approximation. Simulation results for the solvent mediated interactions at the interface differ markedly from the interactions of the particles in the corresponding bulk phases. We find that at short interparticle distances the interactions are considerably more repulsive than those in either bulk phase. At long interparticle distances we find evidence for a long--ranged attraction. We discuss these observations  in terms of interfacial interactions, namely, the three--phase line tension that would operate at short distances, and capillary wave interactions for longer interparticle distances.
\end{abstract}

\pacs{Valid PACS appear here}
\maketitle

\section{\label{sec:level1} Introduction}

Nanoparticles can strongly  adsorb at fluid interfaces, liquid--vapor or liquid--liquid \cite{bresme07}. For nanoparticle sizes between 1 and 10$^2$  nm, the adsorption energy estimated from simple thermodynamics arguments \cite{pieranski80} varies by several orders of magnitude, from  10 to 10$^3$ $k_B T$. It has been shown that the adsorption energy is very sensitive to the nanoparticle geometry \cite{faraudo03} as well as to  interfacial forces such as the line tension, which can affect significantly the stability of the nanoparticles to remain adsorbed at the interface \cite{faraudo03,aveyard96,bresme98,bresme99,bresmepccp}. Moreover, it has been suggested that the coupling between nanoparticle geometry and external fields (electric, magnetic) may cause nanoparticle orientational transitions \cite{bresme07b,bresme08}. 

The stability of nanoparticles at fluid interfaces provides a route to self assemble  two--dimensional arrays and crystals, which have numerous applications in materials science \cite{ozin}. These two--dimensional structures have also attracted the interest of condensed matter researchers, as they provide an opportunity to investigate  two--dimensional phase transitions, either at equilibrium\cite{pieranski80,zahn00,bausch03} or non--equilibrium conditions \cite{zhan97,zhan98,froltsov03}.

The interactions between nanoparticles adsorbed at fluid interfaces are expected to differ from those in the bulk. We have discussed this issue extensively in a recent review article \cite{bresme07}. Nanoparticles move on a fluctuating surface that separates bulk phases (e.g. liquid and vapor), which markedly differ in their permittivities and  densities. These differences in bulk properties are expected to affect the van der Waals interactions, as well as the solvent mediated interactions between particles adsorbed at interfaces. Additionally, fluctuation--induced interactions due to the soft modes of the interface capillary waves may appear, since particles adsorbed at fluid interfaces act as obstacles which modify the interfacial thermal fluctuations. This modification of the capillary wave spectrum can lead  to weak  long--ranged attractions \cite{lehle06,lehle07,lehle08}. Explicit Ising ferromagnet calculations in two and three dimensions have confirmed this notion, showing the existence of attractive long--ranged forces between two points pinned at an interface between two coexisting phases \cite{abraham07}. In addition to the interface fluctuations, single nanoparticle studies suggest that the line tension can modify the wetting behavior of nanoparticles at liquid--vapor interfaces \cite{bresme98,bresme99}. Whether these line tension effects are transferred to the interaction between nanoparticles is not known. The experimental investigation of this question is not easy, given the difficulties in measuring line tension effects at the nanometer scale. Experiments of nanoparticle monolayers could help to tackle these questions and could give important clues on the nature of the interactions operating between nanoparticle adsorbed at fluid interfaces \cite{bera07,lin07,gassin08}.

In order to advance in the understanding of the nature of the interactions operating between nanoparticles at fluid interfaces, we have performed computer simulation and theoretical investigations of the solvent mediated interactions between nanoparticles adsorbed at liquid-vapor interfaces. We have modeled the nanoparticles and the solvent as a simple--liquid mixture interacting with Lennard--Jones (LJ) type potentials. The difference between the solvent--mediated interactions in the bulk phases (liquid and vapour) and those at the interface is of particular interest to us. Since solvent particles of LJ type possess a steep, repulsive core, the solvent--mediated interactions between nanoparticles with a likewise strongly repulsive core should feature  the well--known depletion interactions when their surface--to--surface  separation is less than a solvent diameter \cite{israelachvili}. These depletion interactions play a very important role in determining the phase behavior of polymer--colloid mixtures for instance \cite{meijer91,lekkerkerker92,hansen02}. We expect they may be equally significant in tuning the interactions between nanoparticles adsorbed at fluid interfaces. Unlike the latter,  the depletion interaction of hard--sphere models of nanoparticles and solvent, is very well understood using geometric concepts \cite{Oet09} that extend the well--known Derjaguin approximation to a more appropriate ``colloidal'' limit for the depletion force \cite{Bot09}. For more general solvent models of simple liquids a number of studies exist, see e.g. Refs.~\cite{Shi99,Lou02,Ego04,Arc05}. These results, however, are  not  completely understood theoretically. For larger surface--to--surface distances, i.e. away from the depletion regime, the solvent--mediated interactions between nanoparticles in bulk solvent exhibit oscillations that decay exponentially. These oscillations are connected to packing effects of the solvent particles, and consequently are more pronounced in the bulk liquid that in in the vapor phase. 

In this work we show that the effective interactions between nanoparticles adsorbed at a vapor-liquid interface differ significantly from the interactions in the corresponding bulk phases. These interactions cannot be described in terms of a simple average of the two contributions (liquid and vapor). This finding applies to both the short--range depletion regime, and the regime of intermediate distances (surface--to--surface distances of up to approximately 10 solvent particle diameters). We will give a tentative interpretation of these differences as due to purely interfacial degrees of freedom, namely, the appearance of a three--phase line tension force and interfacial fluctuations. 

The paper is structured as follows: In Sec.~\ref{sec:methods} we introduce the model and briefly describe our theoretical methods (computer simulation and reference--functional integral equations). In Sec.~\ref{sec:results} we present our results for the solvent--mediated interaction between the nanoparticles in the bulk phases and at the interface. The bulk results will be interpreted using the geometrically motivated extension of the Derjaguin approximation whereas for the interpretation of the interface results we introduce the concepts of the three--phase line tension and of fluctuation--induced contributions to the solvent--mediated interactions. Sec.~\ref{sec:conclusions} contains conclusions and an outlook.

\section{Model and methods}
\label{sec:methods}

\subsection{Computer Simulations}
\label{methods-sims}

Molecular Dynamics simulations in the canonical ensemble were performed for a nanoparticle pair at an explicit liquid-vapor interface. We have employed a set up similar to that used by us in previous work \cite{bresme98,bresme99,bresmepccp} . The solvent is modeled as a Lennard-Jones fluid,

\begin{eqnarray}
\label{LJ}
  u_{\rm LJ}(\epsilon,\sigma,r_0,r) &=& 4 \epsilon\left [ \left (\frac{\sigma}{r-r_0}\right)^{12} -
\left (\frac{\sigma}{r-r_0}\right)^{6}\right]  \\  
\label{LJsolvent}
U_{ss} (r) & = & \left\{  \begin{array}{lll} 
  u_{\rm LJ}(\epsilon_{ss},\sigma_s,0,r) - u_{\rm LJ}(\epsilon_{ss},\sigma_s,0,2.5\sigma_s)
   && (r \leq  2.5 \sigma_s)\\ \\
  0 && (r > 2.5 \sigma_s) \end{array} \right. \;,
\end{eqnarray}
\noindent
where we use the interaction strength  $\epsilon_{ss}$ and the solvent diameter $\sigma_s$ to define  dimensionless units for the temperature $T^{*} = k_B T / \epsilon_{ss}$, density $\rho^* = \rho \sigma_s^3$, pressure $p^*=p\sigma_s^3/\epsilon_{ss}$ and surface tension $\gamma^*=\gamma\sigma_s^2/\epsilon_{ss}$. The interactions between the nanoparticle and the solvent are defined in terms of the shifted  Lennard--Jones potential in equation (\ref{LJ}) \cite{bresme98},

\begin{eqnarray}
\label{LJsolvent2}
U_{ns} (r) & = & \left\{  \begin{array}{lll} 
  u_{\rm LJ}(\epsilon_{ns},\sigma_{s},\sigma_{ns}-\sigma_s,r) - 
  u_{\rm LJ}(\epsilon_{ns},\sigma_s,\sigma_{ns}-\sigma_s,3.5\sigma_s)
   && (r \leq  \sigma_{ns}+2.5 \sigma_s)\\ \\
  0 && (r > \sigma_{ns}+2.5 \sigma_s) \end{array} \right. \;, 
\end{eqnarray}
where $\sigma_{ns} = (\sigma_n + \sigma_s) / 2$.  We have chosen $\sigma_n = 7 \sigma_s$ for the nanoparticle diameter. Considering a solvent diameter of about 0.3 nm, this corresponds to a nanoparticle of $\approx$ 2 nm, which is close to typical sizes of small metal passivated nanoparticles \cite{heath97,tay06}. The nanoparticle-solvent interaction was set to $\epsilon_{ns} = 1.5 \epsilon_{ss}$, giving a nanoparticle--interface contact angle of 95$\pm$5 degrees. Because we want to isolate the solvent contributions to the interparticle interactions, the nanoparticle--nanoparticle interactions $U_{nn}(r)$ were set to zero for $r>\sigma_n$, i.e. beyond a hard core. All the simulations were performed for a single thermodynamic state, $T^* = 0.80$ $\approx 0.86 T_c$, where $T_c$ is the critical temperature of the spherically truncated and shifted Lennard-Jones model considered here \cite{alejandre99}. We obtained the following coexistence densities and surface tension for this thermodynamic state: $\rho_{l}^* = 0.732$, $\rho_{v}^* = 0.019$, liquid--vapour surface tension  $\gamma^* = 0.39$, and pressure $p^* = 0.014$. 

The nanoparticles were placed on the surface of a liquid slab consisting of  3$\times 10^4$ solvent atoms and dimensions $\{x,y,z\} = \{41,41,20.5\}\sigma_s$. The interface plane was normal to the $z$ axis. The nanoparticles did not move during the simulations. A typical computation of the force for a specific distance involved long runs, about $5\times10^6$ time steps, with a reduced time step of $\delta t^* = 0.005$, which corresponds in real units to times of the order of tens of nanoseconds. 

The depletion force experienced by the nanoparticles at surface--to--surface separation $d$, is computed through,

\begin{equation}
\label{meanforce}
f(d) = \frac{1}{2}\left< \vect{r}_{ab} \cdot (\vect{f}_{as} - \vect{f}_{bs}) \right>
\end{equation}
\noindent
where $\vect{r}_{ab} = (\vect{r}_a - \vect{r}_b)/|\vect{r}_a - \vect{r}_b|$ is the unit vector along the axis joining the centre of mass of the two particles, $a$ and $b$, and $\vect{f}_{is}$ is the force between the nanoparticle $i$ and the solvent molecules. The brackets represent an ensemble average. Similar computations were performed to obtain the potential of mean force in the bulk phases, vapor and liquid. In these cases the density of the solvent was adjusted to ensure that the density of the fluid far from the nanoparticles corresponded to the coexistence densities reported above. 

In addition to the nanoparticle pair studies, we performed simulations of nanoparticle arrays at the same thermodynamic conditions. We considered a low nanoparticle surface concentration corresponding to an area per nanoparticle of $A / ( N_n  \sigma_s^2) = 240$, where $N_n$ and $A$ are the number of nanoparticles and the interface area respectively. 
Here, we replaced the nanoparticle hard core potential by a short range repulsive contribution, 

\begin{equation}
\label{shortrange}
U_{nn} = 4 \epsilon_{ns} \left (\frac{\sigma_s}{r - r_0} \right)^{12}
\end{equation} 
\noindent
where $r_0 = \sigma_n - \sigma_s$.  The simulations involved typically 20 nanoparticles and $0.5-1\times10^5$ solvent atoms. We used long trajectories, up to 10 ns, to compute the nanoparticle-nanoparticle pair correlation function, $g(r)$, which was later inverted to extract the total potential of mean force, $W_T$, 

\begin{equation}
\label{PMF}
W_T(d) = W(d) + U_{nn}(d) = -k_B T \ln g(d)
\end{equation}
\noindent 
The solvent contribution, $W(d)$, can be  extracted  by subtracting the direct nanoparticle-nanoparticle interaction (equation (\ref{shortrange})) from the total potential of mean force. In these simulations the nanoparticles where free to  fluctuate at the interface.

We also calculated the solvent--mediated force between the nanoparticles immersed in the bulk vapor and liquid state, respectively, as well as for the two additional temperatures $T^*=1.0$ ($p^*=0.80$) and $T^*=1.25$ ($p^*=1.71$) on the isochore  $\rho^* = 0.732$  (i.e. away from coexistence). It is quite instructive to monitor the behaviour of the solvent--mediated force in the depletion region upon approaching coexistence (see below). Since the solvent--mediated force in the depletion region will be interpreted below in the ``colloidal'' limit (large nanoparticle radius), we need the nanoparticle--solvent surface tension $\gamma_{ns}$, which can be obtained from the solvation free energy $F_{np}$ of the nanoparticle by 
\begin{equation}
 \label{eq:gamma_def}
   4\pi R^2\,\gamma_{ns} = F_{np} - \frac{4\pi}{3}R^3\, p\;.  
\end{equation}
This definition of the surface tension is used within the integral equation approach. In order to compare the accuracy of the integral equation approach with simulation data, a different quantity $\tilde\gamma_{ns}$ is evaluated which is easier to simulate, 
\begin{equation}
 \label{eq:gammatilde_def}
  8\pi R\Delta R \,\tilde \gamma_{ns} = \Delta F_{np} - 4\pi R^2 \Delta R\, p\;,
\end{equation}
with $\Delta R$ chosen to be small. In the limit $\Delta R \to 0$, the relation between both surface tensions is given by $\tilde \gamma_{ns} = \gamma_{ns} + (R/2) d\gamma_{ns}/dR$. Note that the values for $\gamma_{ns}$  and $\tilde\gamma_{ns}$ in Table I 
have been obtained using $R=\sigma_{ns}$ for the defining surface. The difference between  $\gamma_{ns}$ and $\tilde\gamma_{ns}$ is small. The differences between  $\gamma_{ns}$ and $\tilde\gamma_{ns}$ should become even smaller as the nanoparticle radius increases. This statement can be made quantitative  by considering a simple cubic form in $R$ for the free energy of insertion, 
\begin{equation}
 \label{eq:finsertion}
F_{np} = 4 \pi \left( p \frac{R^3}{3} + \gamma_\infty R^2 + \kappa R + \bar\kappa\right) \;.
\end{equation}
\noindent
Such a form corresponds to the well--known scaled particle picture, with $\gamma_\infty$ representing the surface tension of a particle in the limit of infinite radius, i.e., a wall, and with $\kappa$, $\bar \kappa$ being correction terms. In the recently developed idea of morphological thermodynamics \cite{Koe04}, the coefficients $\kappa$ and $\bar \kappa$ correspond to the free energy coefficients of the integrated mean and Gaussian curvature of the nanoparticle, and no more coefficients besides $p$, $\gamma_\infty$, $\kappa$, $\bar \kappa$ are needed to describe the nanoparticle solvation free energy. The derivative of equation (\ref{eq:finsertion}) is given by,
\begin{equation}
 \label{eq:finsertion2}
\frac{dF_{np}}{dR} = 4 \pi R^2 \left ( p   + \frac{2 \gamma_\infty}{R} + \frac{\kappa}{R^2} \right )
\end{equation}
\noindent
thus $\tilde \gamma_{ns} = \gamma_\infty + \kappa/(2R)$, which is formally Tolman's equation, showing that the term $(R/2) d\gamma_{ns}/dR$ above decreases as $-\kappa/ (2R)-\bar \kappa/R^2$.
By dimensional arguments, one would expect that $|\kappa\sigma_s/\epsilon_{ss}| \alt 1$ and
 $|\bar \kappa\sigma_s^2/\epsilon_{ss}| \alt 1$, thus for larger nanoparticle radii, $R$, we expect that $\tilde \gamma_{ns}$ and $\gamma_{ns}$ quickly converge. We note that in the limit of very small particles, with diameters a few times the solvent diameter, the main factor influencing the numerical values for the  surface tension is the choice of the defining surface (see ref.~\cite{bresme99} for an illustration of this effect). Although there is some arbitrariness in this choice, it is physically reasonable to locate the surface close to the repulsive core of the nanoparticle--solvent interaction $U_{ns}$, i.e. approximately at the surface of an ``exclusion" sphere around the nanoparticle.
Note that for our choice $R=\sigma_{ns}$ we have $U_{ns}(r=R) =0$, and for smaller values of $r$ the potential quickly rises.

\subsection{Integral Equations}

The thermodynamic properties of the solvent and the solvent--mediated interaction between the nanoparticles are obtained through the pair correlation functions
$g_{ij}(r)=h_{ij}(r)+1$ in the mixture of solvent particles at density $\rho_1 \equiv \rho$  with the nanoparticles at infinite dilution, $\rho_2 \to 0$. Here, species indices $i,j$ are `$s$' for the solvent and `$n$' for the nanoparticles. The first set of relations determining the correlation functions is given by the  Ornstein--Zernike equations. At infinite dilution, they take the following form:
\begin{eqnarray}
  h_{ss}(r) - c_{ss}(r) &=& \rho h_{ss} \ast c_{ss} (r) \;, \\
  h_{sn}(r) - c_{sn}(r) &=& \rho h_{sn} \ast c_{ss} (r) \;, \\
  h_{nn}(r) - c_{nn}(r) &=& \rho h_{sn} \ast c_{sn} (r) \;,
\end{eqnarray}
with $h \ast c(r) = \int d\vect r' h(r) c(|\vect r'-\vect r|)$ denoting the convolution product. The second set, given by the general closure relations
\begin{equation}
 \ln g_{ij}(r) +\beta U_{ij}(r) = h_{ij}(r) - c_{ij}(r) - b_{ij}(r) 
\end{equation}
requires knowledge of the bridge functions $b_{ij}$ ($\beta=1/(k_BT))$. 
We employ the Reference Functional Approximation (RFA) \cite{Oet05,Oet09a} where accurate density functionals ${\cal F^{\rm ref}}[\rho(r)]$ of a hard--sphere reference system are employed to determine the bridge functions:
\begin{eqnarray}
   b_{ij}(r) &=& \beta\left.\frac{ \delta {\cal F}^{\rm B,ref} }{\delta \rho(\vect r)}
                 \right|_{\rho(\vect r) = \rho_j g_{ij}(r)} \;, \\
\label{eq:fbdef}
   \beta {\cal F}^{\rm B,ref}[\rho(\vect r)] &=& \beta({\cal F}^{\rm ref}[\rho(r)] - 
                                                 {\cal F}^{(2),{\rm ref}}[\rho(r)] )\;, \\
    \beta {\cal F}^{(2),{\rm ref}}[\rho(r)] ) &=&
        \beta F^{\rm ex,ref}(\rho_1,\rho_2)-
\sum_{i}\int d\mathbf{r}c_{i}^{(1),{\rm ref}}(\mathbf{r}; \rho_{i} )\Delta \rho _{i}(\mathbf{r})- \\
 \nonumber & &
\frac{1}{2}\sum_{i,j}\int d\mathbf{r}d\mathbf{r^\prime }
c_{ij}^{(2),{\rm ref}}(\mathbf{r,r^\prime};  \rho_i)\Delta \rho_{i}(\mathbf{r})
\Delta \rho _{j}(\mathbf{r^\prime }) \;.
\end{eqnarray}
Here, $\beta {\cal F}^{(2),{\rm ref}}$ is the reference system density functional Taylor--expanded to second order in the density deviations $\Delta \rho_i(\vect r) = 
\rho_i(\vect r) - \rho_i$ around the bulk densities $\rho_1$ and $\rho_2$. Consequently, the first Taylor coefficient, the direct correlation function of first order, is a constant and given by the excess chemical potential of the reference system at bulk density $\rho_i$: $-c_{i}^{(1),{\rm ref}}(\vect r;\rho_i) \equiv \beta \mu_i^{\rm ex,ref}(\rho_i)$. The second Taylor coefficient is the direct correlation function of second order and depends only on the difference of the two position arguments: 
$c_{ij}^{(2),{\rm ref}}(\mathbf{r,r^\prime};  \rho_i) \equiv c_{ij}^{{\rm ref}}(|\vect r-\vect r^\prime|;  \rho_i)$. As can be seen from equation (\ref{eq:fbdef}), the bridge functions are generated from all Taylor coefficients higher than second order in an expansion of the reference free energy functional around the bulk density. The hard--sphere diameter $\sigma_1$ of the reference solvent can be determined via a minimization criterion for the bulk free energy \cite{Oet09a}, for the reference diameter $\sigma_2$ for the nanoparticle we choose $\sigma_2=\sigma_n$.

The RFA usually gives very good results for the equation of state of simple liquids, see Table I 
and refs.~\cite{Oet05,Oet09a}, except for the immediate vicinity of critical points. Furthermore, the accuracy of fundamental measure hard--sphere functionals for highly asymmetric mixtures allows also a reliable determination of nanoparticle--solvent correlations $h_{ns}(r)$. The effective nanoparticle--nanoparticle correlations $h_{nn}(r)$ are related to the solvent mediated potential $W$ by $\beta W(r) = - \ln(h_{nn}(r)+1)$ for $r>\sigma_n$. These correlation functions can be determined to  good accuracy  outside the depletion region ($r>\sigma_n+\sigma_s$) \cite{Amo01,Aya05,Amo07}, inside the depletion region the ``colloidal'' limit expressions of refs.~\cite{Oet09,Bot09} can be employed (see below). The nanoparticle--solvent surface tension has been  evaluated according to equation (\ref{eq:gamma_def}) by noting that the solvation free energy $F_{np}$ of the nanoparticle is the excess chemical potential $\mu_{n}^{\rm ex}$ in the dilute limit which within RFA is given by \cite{Oet05}
\begin{eqnarray}
 \label{eq:muex} 
 \beta \mu_n^{\rm ex}(\rho) &=& 
   \beta \mu_n^{\rm ex, HNC}(\rho) -
   \rho \int d\vect r\; g_{ns}(\vect r)\;
      b_{ns}(\vect r) + 
    \beta \left.{\cal F}^{\rm B}[\rho(\vect r)]\right|_{
    \rho(\vect r)=\rho\; g_{ns}(\vect r)} \;, \\
 \beta \mu_n^{\rm ex, HNC}(\rho) &=& \rho \int d\vect r\;
   \left( \frac{1}{2} h_{ns}(\vect r)\left[
  h_{ns}(\vect r)-c^{(2)}_{ns}(\vect r)\right] - c^{(2)}_{ns}(\vect r) \right) \;.
\end{eqnarray}
By varying the nanoparticle radius, also the surface tension $\tilde \gamma_{ns}$ defined in equation (\ref{eq:gammatilde_def}) could be computed and compared to the corresponding values of the MD simulation (see Table I 
). The agreement between simulated and integral equation results is satisfactory and underlines the usefulness of the RFA approach for very asymmetric mixtures. We verified that the cubic form for $F_{np}$ (equation (\ref{eq:finsertion})) constitutes an excellent fit to the RFA numerical results for nanoparticle radii between 0.5 and 10 $\sigma_s$. In these fits, the pressure $p$ which governs the volume contribution to $F_{np}$ turns out to be almost identical to the virial pressure of the solvent reported in Table I 
. This demonstrates the good thermodynamic consistency of RFA.

\section{Results}
\label{sec:results}

\subsection{Mean force of nanoparticles in bulk phases}

\begin{figure}[!ht]
\centerline{\epsfig{file=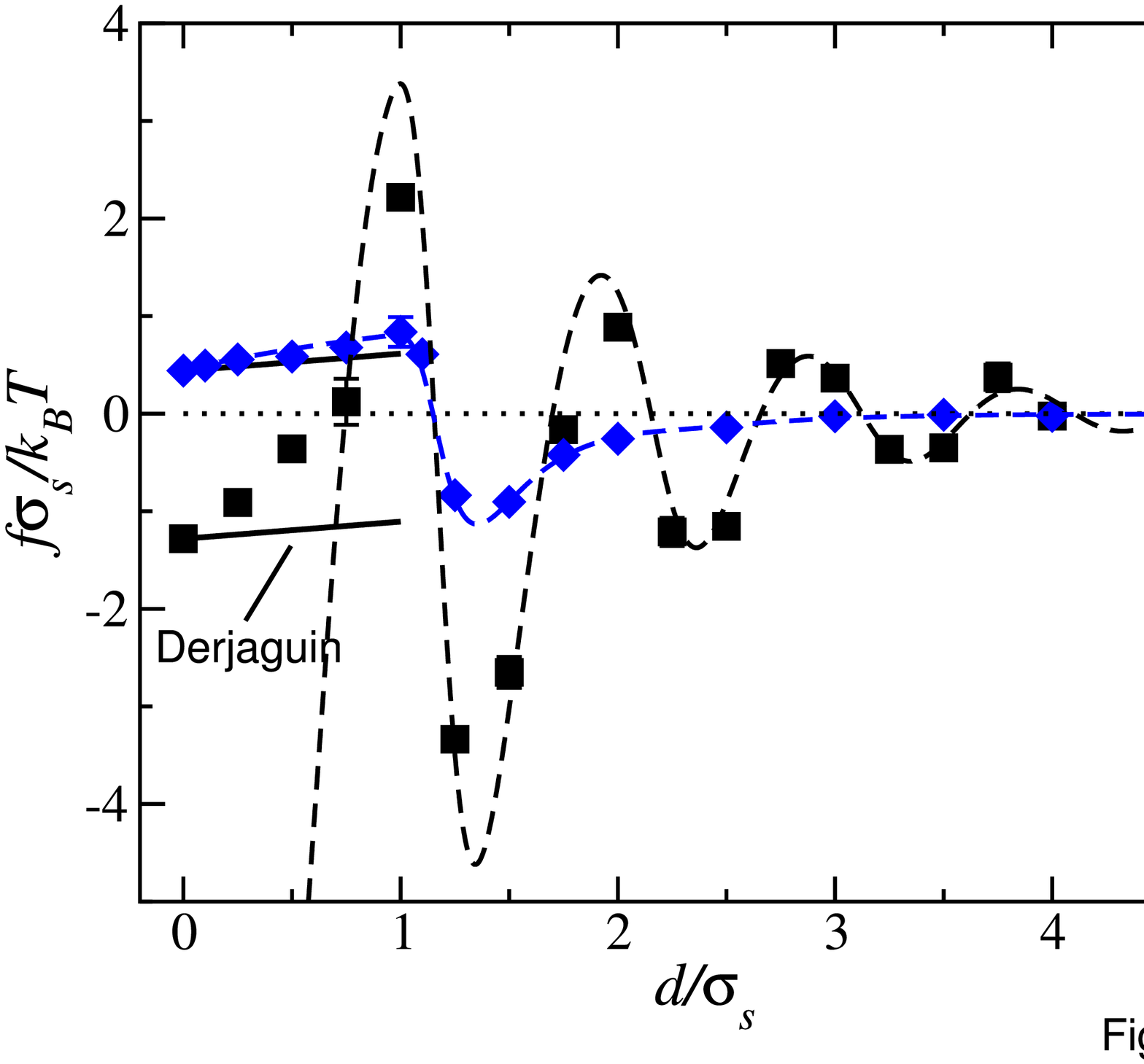,width=10cm}}
\caption{\label{FIG1} Solvation force between two nanoparticles immersed in bulk, vapor (diamonds) and liquid (squares) as a function of the nanoparticle--nanoparticle separation. The dashed lines represent the results from the RFA integral equation discussed in the text. Full lines are the prediction of the Derjaguin approximation. The error bars represent the typical uncertainty associated to the simulation data.}
\end{figure}

Figure \ref{FIG1} shows the mean force $f$ between the nanoparticles as a function of the surface--surface separation $d=r-\sigma_n$ for the bulk phases, liquid and vapor, at temperature  $T^*=0.80$ (coexistence). The MD results for the force in the liquid phase exhibit the characteristic oscillatory behavior expected in a dense fluid, with regular peaks at $\sigma_s$ intervals, and the force is attractive in the depletion region $ d < \sigma_s$. The mean force  in the vapor phase is less structured, as expected, featuring a weak repulsive force in the depletion region, indicating a preference for  the nanoparticles to be solvated in the vapor phase. The mean force is determined in integral equation theory via $\beta f  = -\partial \ln(h_{nn}(r) + 1)/\partial r$. We find that the force between the nanoparticles in the vapor are very well reproduced by the RFA integral equation in the whole interval of distances. This is consistent with previous integral equation studies, which showed that the hypernetted chain (HNC) and Percus--Yevick (PY) integral equations very accurately reproduce the structural and thermodynamics properties of nanoparticles in bulk vapor \cite{bresmejpc}. On the other hand those studies revealed the limitations of these integral equations to predict the properties in the liquid phase.  The RFA integral equation theory employed in this work shows a dramatic improvement with respect to those theories. We find that  the RFA theory reproduced quantitatively the simulation  results for $d > \sigma_s$ (consistent with previous studies \cite{Amo01,Aya05,Amo07}), but fails to capture the magnitude of the attractive force in the depletion region $d<\sigma_s$. 

The mean force between the nanoparticles in the depletion region is governed by the density distribution of the solvent spheres in the annular wedge formed between the nanoparticles. With the radius of the nanoparticle becoming large, the annular wedge induces a quasi--two dimensional confinement for the solvent spheres. Bulk integral equations (as employed here) do not capture this regime well. Explicit density functional studies for hard spheres and large nanoparticle radii \cite{Bot09} suggest the validity of the following expression for the depletion force in the ``colloidal'' limit (large nanoparticle radius):
\begin{eqnarray}
 \label{eq:force_s_morph}
  \frac{f}{\pi R} & \approx & -p x - 2\gamma - \kappa\, \frac{\pi}{2} \sqrt{\frac{1}{Rx}}  \qquad (x= 2R-\sigma_n-d) \;.
\end{eqnarray}
Here, $R$ is an effective radius of the nanoparticle (i.e. an equipotential surface). For hard spheres, the obvious choice would be $R=\sigma_{ns}$, the radius of the exclusion sphere around the nanoparticle. The above expression for the depletion force follows only  from geometric considerations on the overlap region of the two exclusion spheres \cite{Oet09,Bot09}. Let $V(r)$, $A(r)$ and $C(r)$ denote the volume, surface area and integrated mean curvature of this overlap region (which obviously depend on the distance $r$ between the centers of the exclusion spheres). According to morphological thermodynamics \cite{Koe04} which we already briefly introduced in the discussion of the solvation free energy of one nanoparticle (equation (\ref{eq:finsertion})), the depletion potential takes the form $W(r) = -pV(r)-\gamma A(r) - \kappa C(r)$, where $p$ is the pressure in the solvent, $\gamma=\gamma_{ns}$ is the nanoparticle--solvent surface tension and $\kappa$ is the mean curvature coefficient. The mean force follows as $f=-\partial W/\partial r$. Note that with $\kappa=0$, equation (\ref{eq:force_s_morph}) turns into the well--known Derjaguin approximation for the mean force which is widely employed in colloidal physics. The coefficient $\kappa$ can actually be interpreted as a {\em line tension} associated with the edge of the annular wedge between the nanoparticles \cite{martin04}.  Obviously, this line tension becomes important either when $x \to 0$ in equation (\ref{eq:force_s_morph}) or if both $p$ and $\gamma$ are numerically small.

\begin{figure}[!ht]
\centerline{\epsfig{file=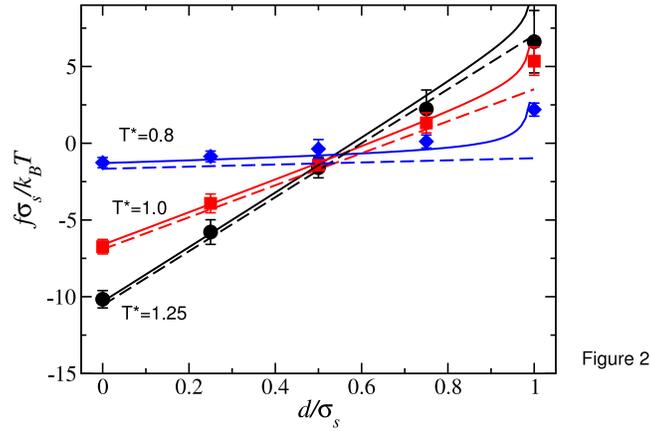,width=10cm}}
\caption{\label{FIG2} Solvation force between nanoparticles immersed in a fluid phase of density $\rho^* = 0.732$.  Symbols represent simulation results for different temperatures, full lines represent the ``colloidal'' limit (equation (\ref{eq:force_s_morph}), with $\kappa^*=-0.03$) and dashed lines are the predictions of the Derjaguin approximation ($\kappa^*=0$).}
\end{figure}

We have tested this form for the mean force on the isochore $\rho^*=0.732$ for the temperatures $T^*=0.80$ (coexistence), 1.0 and 1.25 (see Table I 
). In applying equation (\ref{eq:force_s_morph}), we have used $R=\sigma_{ns}=4\sigma_s$ as one would do for hard spheres 
and we employed the surface tension $\gamma=\gamma_{ns}$ from the RFA at the respective state points (see Table I 
). For the coefficient $\kappa$, we used the value $\kappa^*= \kappa\sigma_s/\epsilon_{ss}=-0.03$ which has been determined by a fit to the simulation results for the depletion force at $T^*=0.80$ and which we use also for the other temperatures.\footnote{In principle, $\kappa$ can be detemined from the insertion free energy of a single nanoparticle with varying radius and thus should be equivalent to the $\kappa$--coefficient appearing in equation (\ref{eq:finsertion}). However, it is not clear how the nanoparticle--solvent potential must be chosen for different $R$ such that the coefficient $\kappa$ is unequivocally determined. This is yet an open question in the application of morphological thermodynamics to soft potentials.} In Figure~\ref{FIG2} the MD data (symbols) are compared to the theoretical expressions (full lines) and the Derjaguin approximation ($\kappa=0$, dashed lines). It can be seen that the Derjaguin approximation is almost on top of the data points for $T^*=1.25$, the state point deep in the liquid but it fails to catch the upturn of the force for $d\to\sigma_s$ for the two lower temperatures. The line tension term indeed can account for this behaviour. It is even the dominant term for the state point right at coexistence where both the reduced pressure $p^*\approx 0.044$ (RFA) and the reduced surface tension $\gamma_{ns}^*=0.031$ (RFA) are numerically small, as anticipated before.

\subsection{Mean force of nanoparticles at the liquid-vapor interface}

\begin{figure}[!ht]
\centerline{\epsfig{file=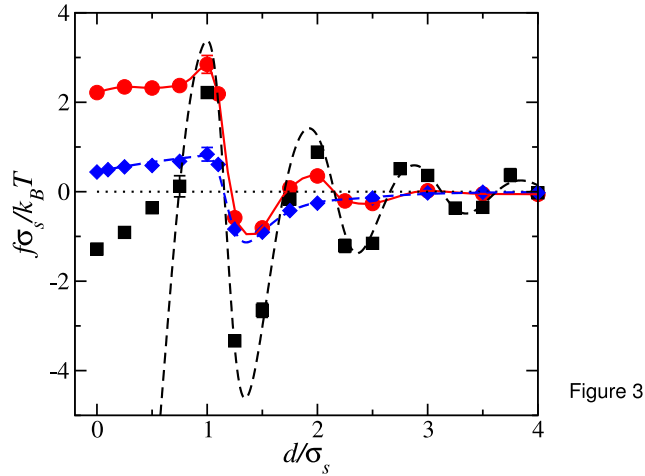,width=10cm}}
\caption{\label{FIG3} Solvation force between nanoparticles adsorbed at the liquid--vapor interface (circles). The line is a guide to the eye. The solvation forces of the nanoparticles in the liquid (squares: simulation results and dashed line: RFA theory) and vapor (diamonds: simulation results and dashed line: RFA theory) phases are also shown for comparison. The error bars represent the typical uncertainty associated to the simulation data.}
\end{figure}

\begin{figure}[!ht]
\centerline{\epsfig{file=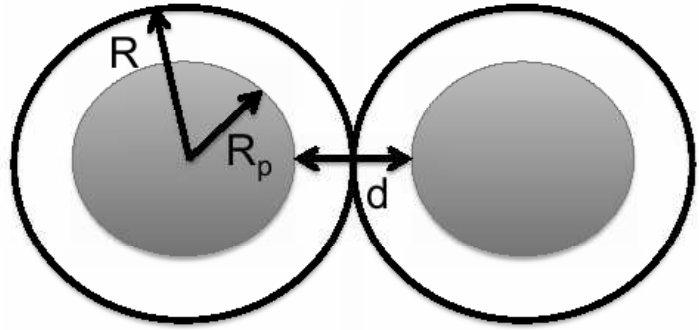,width=10cm}}
\caption{\label{FIG4} Sketch of the geometrical construction used to compute the line tension contribution to the solvation force. $R_p$ is the nanoparticle radius, $R$ the radius of the exclusion sphere and $d$ the surface-surface separation. See text for details.}
\end{figure}

Figure \ref{FIG3} represents the main result of this paper. We report the solvation force between nanoparticles adsorbed at the liquid--vapor interface.  We recall that the solvent--nanoparticle interactions used in our simulations result in a nanoparticle contact angle slightly larger than 90 degrees, showing  there is not a strong preference for the nanoparticles to be immersed in the vapor or liquid phases. Comparison of the force between the nanoparticles at the interface (circles in Figure \ref{FIG3}) with the vapor (diamonds in Figure \ref{FIG3}) and liquid (squares in Figure \ref{FIG3}) counterparts shows that there is not a trivial relationship between the forces at interfaces and in the bulk. This is particularly evident in the depletion region, $d < \sigma_s$, where the interaction between the nanoparticles at the interface is considerably more repulsive than in the bulk phases. The repulsive force, $f^* \approx 2$, amounts to about 8 pN, considering the standard Lennard-Jones parameters for Argon ($\epsilon =0.9962$ kJ/mol, $\sigma = 3.405$\AA). Therefore, the simple geometric description leading to equation (\ref{eq:force_s_morph}) for the force is not valid for the interface case, since it would lead to a weighted superposition of the liquid and vapour forces and thus to an almost zero force in the depletion region. The main conclusion from the data represented in Figure \ref{FIG3} is that the interface adds a contribution to the force that is not present in bulk.  One possible origin for this force could be again a line tension. For nanoparticle separations corresponding to the depletion region, $d < \sigma_s$, there exists a three phase line surrounding the nanoparticle pair. The three phase line will tend to contract for positive line tensions and expand for negative ones, adding an attractive or repulsive contribution respectively, to the total force between the nanoparticles. A repulsive force like the one we observe in our simulations would indicate a negative line tension. We have made an attempt to estimate the order of magnitude of the line tension needed to generate a 
repulsive force of the order of the one observed in the computer simulations. To this end we consider a phenomenological model, whereby the line tension contribution to the free energy is given by, $F_\tau = \tau L$, where  $L$ is the length of the three phase line surrounding the nanoparticle pair (cf. Figure \ref{FIG4}). Hence, within this model the force due to the line tension, assuming a nanoparticle contact angle of 90 degrees  is given by:

\begin{eqnarray}
\label{force-line-tension} 
f_\tau &=& \left(\frac{-\partial F_\tau}{\partial d} \right )_{NVT} = 
\frac{-2 \tau}{\sqrt{1-\left(\frac{ R-x/2}{ R}\right)^2}} 
\stackrel{R\to\infty}{\approx} -2 \tau \sqrt{\frac{R}{x}} \\
\nonumber && \qquad (x= 2R-\sigma_n-d) \;.
\end{eqnarray}
\noindent
This contribution is very similar to the line tension contribution in equation (\ref{eq:force_s_morph}) for the depletion force in the bulk. $R$ is the radius defining the three phase line (see Figure  \ref{FIG4}). A natural choice for this radius would be $R = \sigma_{ns}$, as before, but one could equally well choose a
somewhat larger value (e.g., where the nanoparticle--solvent potential $U_{ns}$ is minimal). Because at this molecular scale it is difficult to split up the force contributions due to part of the nanoparticle being in the vapor phase and other part in the liquid phase, we can only  attempt to estimate the sign and order of magnitude of the line tension. Considering equation (\ref{force-line-tension}), the magnitude of the force when the nanoparticles are in contact ($d=0$), and 
$R = \sigma_{ns}$, we estimate a line tension of the order of  $\tau^* = -0.5$.  The order of magnitude and sign of the line tension obtained from this analysis is similar to the 
values reported  from computations of nanoparticles adsorbed at Lennard--Jones liquid--vapor interfaces, where line tensions of the order of $\tau^*  \approx -0.3$ were found \cite{bresme98,bresme99}. 

\begin{figure}[!ht]
\centerline{\epsfig{file=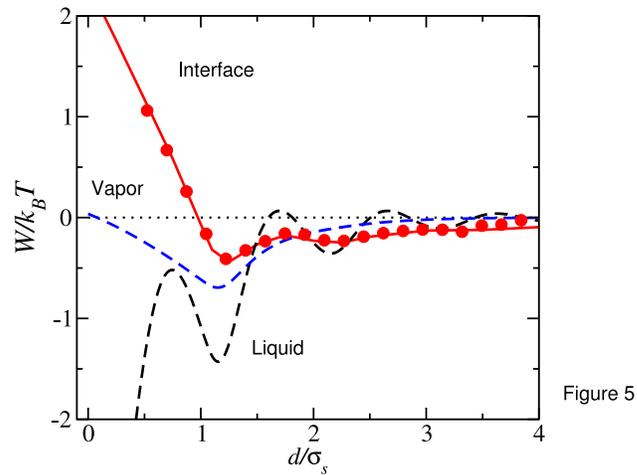,width=10cm}}
\caption{\label{FIG5} Potential of mean force for the nanoparticle pair in bulk and at the interface. Dashed lines represent the results from the RFA  integral equation. Full lines represent the potential of mean force of the nanoparticle pair at the interface obtained from computer simulations. The potential of mean force for a nanoparticle array obtained from inversion of the nanoparticle pair correlation function is also shown (circles).}
\end{figure}

To further test the existence of the extra repulsive contribution  between the nanoparticles in the depletion region, we have performed simulations of nanoparticle arrays adsorbed at the liquid-vapor interface (see section \ref{methods-sims} for details). Figure \ref{FIG5} shows the corresponding potentials of mean force (PMF), which were obtained from integration of the forces for the nanoparticle pair, and from inversion of the pair correlation function (equation (\ref{PMF})) for the nanoparticle array.  The agreement between both computations in the depletion region is excellent, showing that the PMF obtained from the nanoparticle pair represents an accurate approximation for the short range interactions between particles at the low coverages investigated in this work.  Comparison of the 
PMF at the interface with the bulk phases (c.f. Figure \ref{FIG5}) makes even clearer the strong impact that the interface has on the nanoparticle interactions.

\begin{figure}[!ht]
\centerline{\epsfig{file=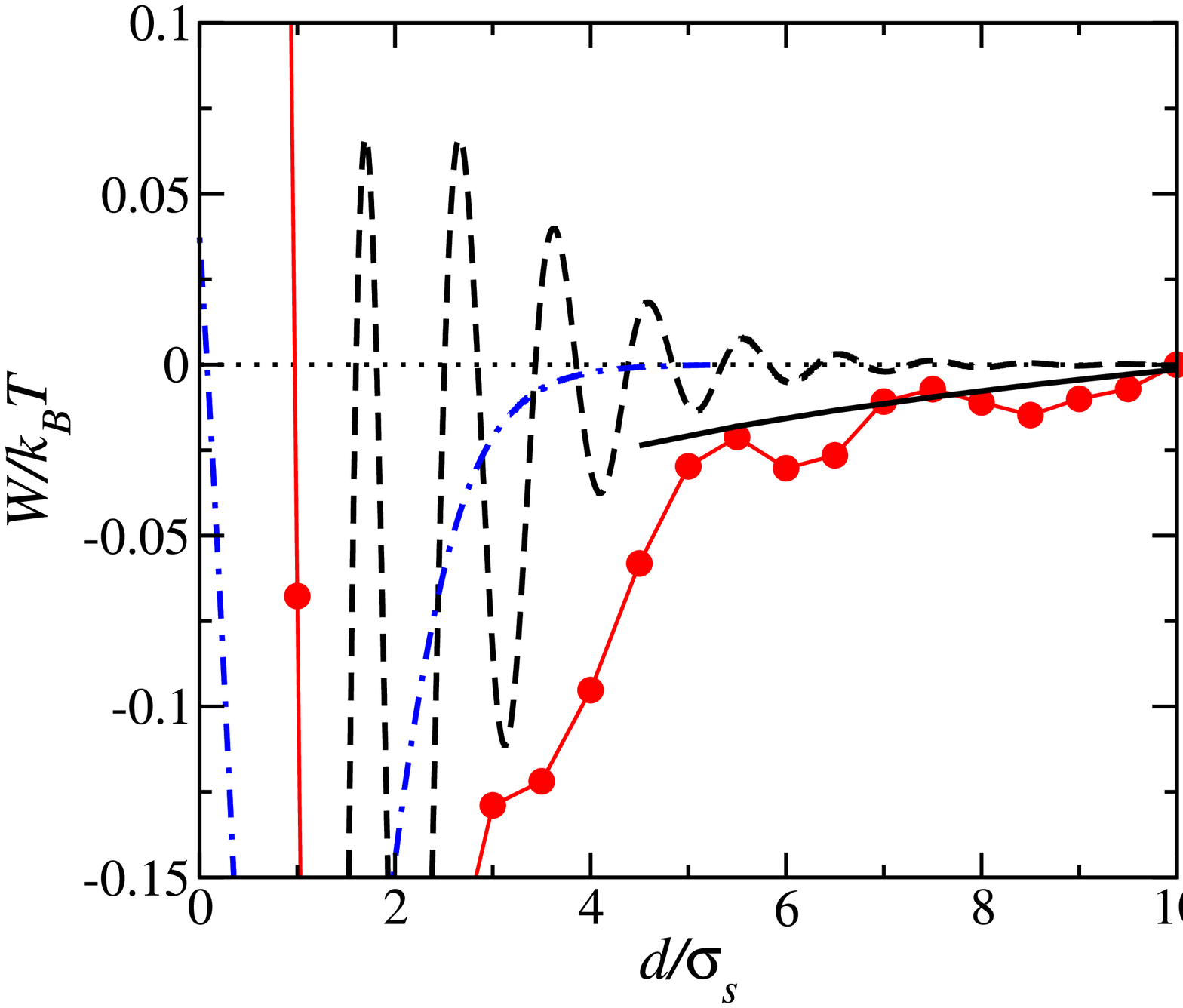,width=10cm}}
\caption{\label{FIG6} The long range behavior of the potential of mean force: liquid (dashed line: RFA  integral equation), vapor (dashed-dotted line: RFA  integral equation), interface  (circles and thin line: simulation results), and theoretical prediction from the fluctuation force equation (\protect\ref{eq:flucfull})  (full line).}
\end{figure}

Finally, we discuss the long range behavior of the potential of mean force between nanoparticles at the liquid--vapor interface (c.f. Figure \ref{FIG6}). For the purpose of analyzing the decay of the interactions in the bulk phases we have considered the integral equation results. We showed above that the RFA integral equation provides an excellent approximation to the simulation results. The potential of mean force for the bulk phases, liquid and vapor,  decays to zero for nanoparticle separations being of the order of 6$\sigma_s$, whereas it shows a  longer--ranged behavior in the case of  nanoparticles adsorbed at the interface.  
(Note that in the integration of the MD force data at the interface, we have set the potential to zero for the longest distance investigated in this work (10$\sigma_s$)).
The occurence of a long--ranged tail in the PMF at the interface might be connected with capillary waves. It has been suggested that colloids adsorbed at fluid interfaces act as obstacles which
 perturb the fluctuations of the capillary waves through boundary conditions at the three phase contact line \cite{lehle06}. For colloids placed at the interface at center--to--center distance $r$, the fluctuation spectrum and consequently the free energy of the capillary waves will depend on $r$. This results in a distance--dependent fluctuation force, which has been described as a variant of the Casimir force (originally thought of as induced by quantum fluctuations). It has been shown that the fluctuation force is strongly dependent on the boundary conditions applying to the colloids, namely, whether the colloids are fixed or fluctuate. For fixed colloids like the ones investigated here and a freely fluctuating three--phase contact line, the fluctuation force in the limit of the capillary length $\lambda \to \infty$ is given  by \cite{lehle06,lehle07, lehle08},

\begin{equation}
\label{fluctuation-force}
\beta U_{fluc} \approx \frac{1}{2} \ln \left( 1+ \ln \frac{2r}{\sigma_n} \right ) + {\rm const.}  + {\cal O}\left(\frac{\sigma_n^4}{r^4}\right)\;.
\end{equation}
\noindent
This equation was derived in the context of a structureless solvent, where the spectrum of the interface position fluctuation is entirely governed by the capillary wave hamiltonian \cite{Buf65}). Thereby molecular effects are ignored. We expect that equation (\ref{fluctuation-force}) will become a good  approximation for long interparticle distances, since in this case the molecular nature of the solvent becomes secondary with respect to the fluctuations of the interface. Nonetheless, it is difficult to quantify the distance at which molecular effects become unimportant.  This distance can be estimated considering the results reported in a very recent work. It has been shown that the dynamics of nanoscopic capillary waves on Lennard--Jones liquid surfaces shows very good agreement with the hydrodynamic theory down to very small wavelengths, of about four molecular diameters \cite{delgado08}. This result indicates that a continuum approximation may provide an accurate approach at  very small length scales. 

We note that the finite size of our simulation box sets a cutoff for the capillary wave spectrum. Hence, equation (\ref{fluctuation-force}) can not be used directly. The full leading term in $U_{fluc}$ for a finite capillary length in an expansion in $\sigma_n/r$ is given by  \cite{lehle07,lehle08}
\begin{equation}
 \label{eq:flucfull}
 \beta  U_{fluc} =  \frac{1}{2} \ln \left( \left( -\ln\frac{\sigma_n}{2\Lambda} +1 \right)^2 + 
\ln^2\frac{r}{\Lambda}   \right) + {\cal O}\left(\frac{\sigma_n^4}{r^4}\right)\;,
\end{equation} 
where $\Lambda \approx 1.12 \lambda$. In Figure \ref{FIG6} we show the predicted  fluctuation force  for a capillary length $\lambda=20\,\sigma_s$ which corresponds roughly to half the lateral box size used in the simulations and is thus intuitively the correct cutoff for the capillary waves. (Note that also $U_{fluc}$ has been set to zero at $d=10\,\sigma_s$ which corresponds to $r=17\,\sigma_s$). We find that this force has an order of magnitude and decay that agrees well with the simulation results, indicating that capillary fluctuations could add a non negligible contribution to the interactions between nanoparticles at long separations. The small magnitude of $U_{fluc}$ is also due to the smallness of the capillary length; for a realistic capillary length of about 1 mm, the fluctuation--induced potential increases by a factor 4 \dots 5 in the distance regime shown in Figure \ref{FIG6}.

\section{Summary and final remarks}

\label{sec:conclusions}

We have reported computer simulations and integral equation studies of the interactions between a nanoparticle pair immersed in bulk liquid or vapor and  adsorbed at the liquid--vapor interface of  a simple liquid. We have considered a situation in which the nanoparticles have a contact angle of the order of 90$^o$, hence they do not show a strong preference for the vapor or the liquid phase. 

The potential of mean force for the nanoparticles in the bulk phases can be explained well by a combination of geometric arguments for the depletion region (where
the surface--to--surface separation of the nanoparticles is less than one solvent diameter) and of reference functional integral equations for larger separations.
We have found that the potential of mean force between the nanoparticles adsorbed at the interface is significantly different from that obtained in the liquid and vapor phases.  The interactions at the interface cannot be explained as a simple average of the interactions in the corresponding bulk phases. Specifically,  in the depletion region the nanoparticles  repel each other more strongly than in any of the bulk phases. This repulsive interaction is of the order of 2--3 $k_{B} T$  at zero nanoparticle separation. We have confirmed the existence of this repulsive interaction  using computer simulations of low density nanoparticle monolayers adsorbed at the liquid--vapor interface, where the potential of mean force can be alternatively obtained from inversion of the nanoparticle--nanoparticle pair correlation function.

Our analysis suggests  that the line tension associated to the three phase line surrounding the nanoparticle pair may add a significant contribution to the nanoparticle interactions in the depletion region. We have estimated that the line tension needed to induce a repulsive force of the order of the one observed in the simulations is negative and of the order of $\approx-5\times10^{-12}$N. This order of magnitude and sign agrees with  the line tensions estimated in previous investigations of nanoparticles adsorbed at liquid--vapor interfaces \cite{bresme98,bresme99}. We have also shown that the interactions between the nanoparticles 
immersed in the bulk phases (vapor and liquid) decay to zero within about six solvent molecular diameters, whereas the interaction between nanoparticles at the liquid--vapor interface features a longer range attraction that appears to extend beyond ten solvent molecular diameters. The magnitude and decay of the attractive potential is of the order predicted from a theoretical analysis of the force arising from the perturbation of the interfacial capillary fluctuations  \cite{lehle06,lehle07}. Our work suggests that interfacial (line tensions and capillary waves) degrees of freedom may add a non negligible contribution to the depletion forces between 
nanoparticles at interfaces. Overall, our model of nanoparticles adsorbed at  realistic liquid--vapor interfaces  indicate that  the depletion force contributes with 
$\approx$ 1--2 $k_B T$ to the total interaction between nanoparticles at interfaces.

\begin{acknowledgments}
FB would like to thank The Royal Society for financial support and the Imperial College High Performance Computing Service for providing computational resources. MO thanks the German Science Foundation (DFG) for financial support through the  Collaborative Research Centre (SFB--TR6) ``Colloids in External Fields", {project N01}.
\end{acknowledgments}

\bibliography{paper}

\newpage

\begin{table}
  \begin{tabular}{llllllll}
  \hline 
    & & \multicolumn{2}{c}{MD} & \phantom{aa} & \multicolumn{3}{c}{RFA}  \\
    \cline{3-4} \cline{6-8}
       $\rho^*$ & $T^*$  & $p^*$ & $\tilde\gamma_{ns}^*$   & &$p^*$  & 
$\tilde\gamma_{ns}^*$ & $\gamma_{ns}^*$  \\
      \hline 
       0.019\phantom{$-$} & 0.80   &       0.014 &                                        & \phantom{aa} & 0.013    & & {$-$}0.027                     \\
       0.732                          & 0.80   &       0.014 &    \phantom{$-$}0.11   &  \phantom{aa} &0.044    & \phantom{$-$}0.12           & \phantom{$-$}0.031  \\ \hline 
       0.732                          & 1.0     &       0.80   &    $-$0.02                      &  \phantom{aa} &0.83       &  $-$0.05                             & $-$0.14                   \\
       0.732                          & 1.25   &       1.71   &    $-$0.21                      & \phantom{aa}  &1.75       & $-$0.27                              & $-$0.35    \\ \hline
  \end{tabular}
\caption{The thermodynamic states of the cut--off and shifted LJ solvent considered in this work. The nanoparticle--solvent surface tension $\gamma_{ns}$ is defined in equation 
(\ref{eq:gamma_def}), while the modified surface tension $\tilde \gamma_{ns}$ is defined in  equation (\ref{eq:gammatilde_def}). ``MD'' refers to results from molecular dynamics simulation, and ``RFA'' stands for integral equations in the reference functional approximation.}
\label{states}
\end{table}

\newpage

\end{document}